\documentclass[aps,prl,preprint,nopacs,superscriptaddress]{revtex4}

\usepackage{graphicx}
\usepackage{verbatim}
\usepackage{mathrsfs}
\usepackage{gensymb}
\usepackage{color}
\usepackage{ulem}

\usepackage{amsmath,amsfonts,amssymb}
\usepackage{graphicx}

\def\3{2.8in}    
\def\2{2.5in}
\def\4{3.0in}\def \beq {\begin{equation}}
\def \eeq {\end{equation}}
\begin{document}

\title{Band Topology and Dichroic Signature of Bismuth}

\author{Guang~Bian}
\affiliation {Department of Physics and Astronomy, University of Missouri, Columbia, Missouri 65211, USA}

\author{Tay-Rong~Chang}
\affiliation {Department of Physics, National Cheng Kung University, Tainan 701, Taiwan}

\author{Xiaoxiong~Wang}
\affiliation {College of Science, Nanjing University of Science and Technology, Nanjing 210094, China}

\author{Hsin~Lin}
\affiliation {Institute of Physics, Academia Sinica, Taipei 11529, Taiwan}

\author{T. Miller}
\affiliation {Department of Physics, University of Illinois at Urbana-Champaign, 1110 West Green Street, Urbana, Illinois 61801-3080, USA}
\affiliation {Frederick Seitz Materials Research Laboratory, University of Illinois at Urbana-Champaign, 104 South Goodwin Avenue, Urbana, Illinois 61801-2902, USA}

\author{T.-C.~Chiang}
\affiliation {Department of Physics, University of Illinois at Urbana-Champaign, 1110 West Green Street, Urbana, Illinois 61801-3080, USA}
\affiliation {Frederick Seitz Materials Research Laboratory, University of Illinois at Urbana-Champaign, 104 South Goodwin Avenue, Urbana, Illinois 61801-2902, USA}


\date{\today}

\begin{abstract}

Bismuth has been the key element in the discovery and development of topological insulator materials. Previous theoretical studies indicated that Bi is topologically trivial and it can transform into the topological phase by alloying with Sb. However, recent high-resolution angle-resolved photoemission spectroscopy (ARPES) measurements strongly suggested a topological band structure in pure Bi. To address this issue, we study the band structure of Bi and Sb films by ARPES and first-principles calculations. By tuning tight binding parameters, we show that Bi quantum films in topologically trivial and nontrivial phases response differently to surface perturbations. Therefore, we establish an experimental criterion for detecting the band topology of Bi by spectroscopic methods. In addition, our circular dichroic photoemission illuminates the rich surface states and complex spin texture of  the Bi(111) surface. 

\end{abstract}

\maketitle

\section{Introduction}
Bi and Sb have played the pivotal role in designing and developing topological condensed matter phases with protected electronic states, which can be largely attributed to their unique electronic properties, especially the strong spin-orbit coupling \cite{Hofmann, Koroteev, Bian4, Wang, Kimura, Aguilera}. The electronic structure of Bi and Sb has be measured in several experiments by angle-resolved photoemission spectroscopy (ARPES) \cite{Hirahara, Kobayashi, Jia1, Takayama, Hirahara1, Sasaki, Ast, Simon, Hirahara2}. The first-ever three dimensional topological insulator (TI) has been realized in the Bi$_{1-x}$Sb$_{x}$ alloy \cite{TI1, TI2, Review, Hsin_RMP}. Both Bi and Sb are semimetals with a negative band gap which separates the conduction and valence bands everywhere in the Brillouin zone. This separation of the conduction and valence bands enables a definition of $\mathbb{Z}_{2}$ topological invariant for Bi and Sb. Sb has been identified to be a semimetal with the same band topology as the alloy TI, whereas Bi doesn't belong to the topological phase according to previous theoretical studies \cite{Fu3D, Teo}. The topological distinction between the two elements originates from a band inversion at the $L$ point. The two low-energy conduction and valence bands possess opposite parity eigenvalues at $L$, and thus a band inversion leads to a change in the band topology \cite{LiuAllen}. The closure of the $L$ gap has been reported in early transport experiments on Bi-Sb alloys \cite{Lenoir, LLi}. However, this experimental result is not enough to determine the band topology of Bi and Sb, because the band topology of Bi and Sb depends on the band ordering around the $L$ gap. Recent high-resolution ARPES experiments examined the bulk and surface bands of Bi and strongly suggested a nontrivial band topology in pure Bi \cite{Ohtsubo1, Ohtsubo2, Matsuda}. If Bi and Sb were both topologically nontrivial, then there would be no need to close the band gap for the realization of the alloy TI. This is a fundamental issue related to the parent compounds of the very first TI. The decisive evidence for Bi band topology is the parity eigenvalues of the bulk bands at $L$, however this has not yet been conclusively probed in spectroscopic measurements  \cite{Ohtsubo1}. Therefore, there is a pressing need for a practical route to determining the band topology of Bi.


In this work, we study the band structure of both Bi and Sb by a combined method of molecular beam epitaxy (MBE), APRES and first-principles calculations. Our photoemission measurements show that Sb is topological in  film forms. For small film thicknesses ($\le$~21 atomic layers), the surface band of Bi thin films connect to the bulk bands in the same way as in Sb films. Our simulations, however, show that the connection pattern of surface bands with bulk quantum well states alone cannot pin down the true band topology of Bi, because the quantum confinement in thin films can open a hybridization gap at $L$. Interestingly, the surface band dispersion of topological and non-topological films response to surface perturbations in dramatically different ways. The difference in surface band behaviors of thin films roots in the topological property of bulk bands and this feature can be straightforwardly examined by spectroscopic techniques such as ARPES. Therefore, we establish an experimental criterion for directly detecting the topology invariant of pure Bi. Moreover, Bi has been considered an ideal spintronics material for its unique spin texture, small Fermi wavevector, and rich surface electronic states. We perform circular dichroism photoemission, a tool sensitive to spin-polarized surface states, on Bi(111) films. The results unambiguously identifies these essential ingredients of the Bi(111) surface for spintronics applications.


\section{Experimental and Computational Methods}
For the sample growth, an $n$-type Si(111) wafer (Sb-doped with a resistivity $\sim$ 0.01~$\Omega\cdot$cm) was used as the substrate. It was cleaned by direct current heating to yield a 7$\times$7 reconstructed surface \cite{Bian1, Yaginuma1}. Bi and Sb were deposited onto the substrate maintained at 60 K. For Sb films, a 1/3-layer Bi was introduced onto the Si substrate before the deposition of Sb to ensure the smoothness of epitaxial Sb films \cite{Bian2, Bian3}. The amount of deposition was measured by a quartz thickness monitor. The film after deposition was disordered based on in situ electron diffraction. Annealing was performed by passing a current through the sample to optimize the structural quality of films. Photoemission measurements of the band structure were performed at the Synchrotron Radiation Center using 22 eV photons and a Scienta SES-100 analyzer. The energy and momentum resolutions were 15 meV and 0.01 $\text{\AA}^{-1}$, respectively. The base pressure of the ARPES measurement was better than 10$^{-10}$ Torr. 

First-principles band structure calculations were performed using the projector augmented wave method \cite{PAW} as implemented in the VASP package \cite{VASP} within the generalized gradient approximation (GGA) schemes. 11$\times$11$\times$11 and 11$\times$11$\times$1 Monkhorst-Pack $k$-point meshes with an energy cutoff of 400 eV were used in the bulk and slab calculations, respectively. The spin-orbit coupling effects were included self-consistently. To calculate the surface electronic structures with a semi-infinite slab, we first constructed a tight-binding Hamilton for Bi and Sb from first-principles results. The tight-binding matrix elements were calculated by the projection of Wannier orbitals \cite{Wannier} which were obtained by the VASP2WANNIER90 package \cite{Wannier1}. The surface states were then calculated from the surface Green's function of the semi-infinite system \cite{HJZhang}. 

\section{Band structure and topology of Bismuth and Antimony}
Bulk Bi and Sb crystalize in the rhombohedral structure shown in Fig.~1(a). The lattice is composed of two interpenetrating, diagonally distorted face-centered-cubic (FCC) sublattices with two atoms per unit cell. The lattice constants in our calculations were adopted from Ref. \cite{LiuAllen}. The bulk Brillouin zone of the rhombohedral lattice is given in Fig.~1(b) with high-symmetry momentum points marked explicitly. The crystal structure can also be viewed as hexagonal atomic layers stacked along the (111) direction. The interlayer interaction is much weaker than the covalent bonds within each layer, making (111) films behave like van der Waals materials. This picture is confirmed by the fact that Bi is brittle and easily cleaves along the (111) plane. The projected (111) surface Brillouin zone is plotted in Fig.~1(b). Both Bi and Sb  are semimetals with a small overlap between the valence and conduction bands (see Fig.~1(c)). The standard PBE-type GGA pseudopotentials were employed in our band structure calculations. The bulk band structure of Bi and Sb from our simulations is consistent with previous tight-binding and first-principles results. The dominant feature of Bi and Sb bulk band structures close to the Fermi level is the small electron and hole pockets on the Fermi surface, which determine the low-energy electron dynamics of these materials. The electron pocket is located at point $L$ (for both Bi and Sb) while the hole pocket lies at point $T$ (for Bi) or a non-high-symmetry point (for Sb). Interestingly, there is a continuous band gap traversing the whole Brillouin zone. In other words, the conduction and valence bands are separated by this band gap everywhere in $k$ space, which allows an unambiguous definition of the $\mathbb{Z}_{2}$ topological invariant in the same way as the $\mathbb{Z}_{2}$ invariant of topological insulators \cite{Fu3D}. The crystal lattice of Bi and Sb are centrosymmetric, therefore the topological invariant can be determined by examining the parity eigenvalues of valence states at time-reversal invariant momentum (TRIM) points \cite{Fu3D}. We found that the $\mathbb{Z}_{2}$ topological index of Sb is 1 whereas the $\mathbb{Z}_{2}$ number of Bi is zero. This agrees with previous theoretical works \cite{Fu3D, Teo, Bian2,  HJZhang}. To view this topological distinction between Sb and Bi intuitively, we also calculated the Wannier charge center evolution for the time-reversal invariant planes \cite{Yu2011}. The result is displayed in Fig.~1(e). It shows that the $\mathbb{Z}_{2}$ invariant $\nu$ is 0 (the trajectory of Wannier center is made of closed loops) in $k_{z}=0$ plane for both materials while $\nu =$ 1 (the trajectory of Wannier center is an open curve traversing the whole Brillouin zone) and 0 in $k_{z}=\frac{\pi}{2}$ plane for Sb and Bi, respectively, which indicates that Sb is a ``strong" topological material while Bi is not. The change in the band topology is brought about by a band inversion at the $L$ point \cite{Fu3D, Teo}.  We note that thought first-principles calculations indicate that Bi is topologically trivial, there is no conclusive experimental evidence for this theoretical conclusion. The critical information is the parity eigenvalues of the two bulk bands around the small $L$ gap ($\sim$ 0.012 eV \cite{LiuAllen, Aguilera}). It has also been shown that slightly tuning the tight binding parameters can retain the same bulk band dispersion but opposite parity ordering at $L$ \cite{Ohtsubo1}. Here we examine the band structure of Bi bulk and films in both scenarios (topological and non-topological) and identify experimental signatures that can prove the band topology of Bi.


\section{ARPES results and Calculated bands of Thin Films}
Recent high-resolution ARPES measurements on Bi showed that the surface bands of Bi connect the conduction and valence bulk bands in the same fashion as the surface bands of a topological insulator \cite{Ohtsubo1, Ohtsubo2, Matsuda}. This connection leads to an odd number of Fermi surface contours, implying a nontrivial band topology of Bi. To visualize this band connection pattern, we performed ARPES measurements on Bi and Sb films that were grown by molecular beam epitaxy (MBE) on Si(111) substrates. The results from 20L Sb(111) and 20L Bi(111) films are shown in Fig.~2(a) and 2(b), respectively. In the Sb spectrum, there is a Rashba-like surface band around the zone center ($\overline{\Gamma}$) right below the Fermi level. The bulk bands are quantized into quantum well subbands due to the intrinsic quantum confinement effect of thin films \cite{TC}. At $\overline{M}$, there are one conduction subband and a series of valence subbands, which are separated by an energy gap. Therefore, it is straightforward to determine how the surface band is connected to the conduction and valence bulk bands by counting the quantized bands at $\overline{\Gamma}$ and $\overline{M}$, two TRIM points. One surface state branch is connected to the lowest conduction band (C1) and the other to the highest valence band (V1), as schematically shown in the zoom-in spectrum in Fig.~2(C). This corresponds to a nontrivial connection which can be adiabatically transformed into a surface Dirac cone, and therefore signifies the nonzero topological band index of Sb. The surface bands in the Bi film spectrum behaves in a similar way as their counterparts in the Sb film spectrum except that the surface bands overlap the bulk band at $\overline{\Gamma}$. This makes two surface state branches emanate from the convex bulk band dome near $\overline{\Gamma}$ rather than reside inside the band gap. Nevertheless one can easily find that this connection between the surface bands and the bulk subbands would lead to a Dirac cone-like surface band signature if the conduction band minimum at L could be adiabatically elevated above the Fermi level; see the schematic in Fig. 2(d). The film ARPES results are consistent with our first-principles slab calculations. The shaded area in Figs. 2(e, f) gives the projected bulk band region from which one can clearly see how the surface band disperses and evolves into bulk subbands. The results confirm the connection pattern observed in the ARPES experiment, suggesting Sb and Bi films shares the same band topology. However, in our bulk band calculations, we have shown that Sb and Bi have different topological invariants. There seems a conflict between the bulk and film results. We will show next that the quantum confinement effects can reconcile these seemingly contradicting results and provide a route to determining the true band topology of Bi in a spectroscopic measurement.

\section{How to reveal bulk band topology in thin films}
In thin films, the tunneling effect between the two surfaces can open gaps in surface bands at TRIM points \cite{Bian3, Xue1} and consequently alter the connection of the surface bands to the bulk bands. To examine the connectivity of surface bands of Bi and Sb, we calculated the band structure under the bulk limit. The band spectra of semi-infinite crystals with a (111) surface are plotted in Figs.~3(a) and 3(b) for Sb and Bi, respectively. In the Sb case, the two surface branches merge into conduction and valence bands separately as the bands disperse from $\overline{\Gamma}$ and $\overline{M}$, showing the same connection as the film case. By contrast, the surface bands of semi-infinite Bi crystal behave differently as the two surface branches converge and become degenerate at $\overline{M}$, forming a surface Dirac cone. Considering the fact that one $\overline{\Gamma}$ and three $\overline{M}$ form the set of TRIMs for the (111) surface Brillouin zone of Bi, there are even number Fermi surface contours (from surface states) that surround the TRIMs. The semi-infinite result indicates a trivial band topology of Bi, consistent with the bulk band result. Therefore, from a comparison between finite thickness and semi-infinite cases, we find that the apparent nontrivial band topology of Bi(111) thin films is an emergent property that is induced by the quantum confinement effects in thin films. The tunneling effect can be effectively removed by applying a surface perturbation that breaks the degeneracy of the surface states on the two opposite faces. To uncover the band topology of pure Bi in our slab simulations, we introduce a potential bias to the top surface layer of a 20L Bi(111) film. The resulting bands in Figs.~3(c) indeed show the disappearance of the hybridization gap at $\overline{M}$. Now there are two copies of the Dirac cone at $\overline{M}$ with an energy shift, which corresponds to the surface states on the two opposite surfaces. If we look at only the top surface, the two surface bands (the bright curves above the Fermi level in the projected band spectrum in Fig.~3(d)) become degenerate at both $\overline{\Gamma}$ and $\overline{M}$ and give the same band topology as the semi-infinite case. Generally, the hybridization between the states on the two opposite surfaces diminishes as the film thickness increases. This is shown in the band spectrum of films with different thicknesses in Figs.~3(e-f).  The monotonic decrease of the gap can also be seen in Fig.~3(g) which presents the size of hybridization gap at $\overline{M}$ as a function of film thickness. The bulk band gap at $L$ of bismuth is very small ($\sim$ 0.1 eV from our GGA calculations), therefore, the hybridization gap at $\overline{M}$ exceeds the bulk gap for very thin films and thus pushes the surface bands into bulk band region. This explains the observed nontrivial surface band connection with the bulk bands in the ARPES spectra of Bi thin films. In the 42L case, the hybridization gap is smaller than the bulk band gap, and thus at least one of the surface bands has to stay within the bulk gap. We can define a critical thickness at which the hybridization gap is equal to the bulk gap. For Bi, the critical thickness is 21L according to our calculations. We note that the experimental value of the gap size is smaller ($\sim$ 0.012 eV), so the critical thickness for the real material should be larger than the theoretical value. Only beyond the critical thickness can the electronic structure of freestanding films reveals the band topology of the bulk material. In Sb films, the hybridization gap is always larger the the bulk gap as a consequence of topological constraint; see Fig.~3(h). Therefore, the thickness dependence of the film hybridization gap can be used to distinguish between different bulk band topologies. 
%

The surface states of a topological insulator are protected by the time reversal symmetry and they are robust against the presence of non-magnetic impurities and other defects. By contrast, the surface states of a conventional insulator can be eliminated by perturbations on the surface. The robustness of the surface band is, therefore, a defining characteristic of topological materials. In our simulations, Sb and Bi bulks are in different topological phases, so their surface states should behave differently in the presence of surface perturbations. The band structures of semi-infinite Bi and Sb under different surface potential bias $V_{surface}$ are plotted in Fig.~4. For small potential bias (e.g. $V_{surface}=\pm 0.6$ eV), the surface bands of Bi are shifted in energy but they still reside inside the bulk band gap. This surface bands gradually move out of the bulk gap as the potential bias further increases as shown in Figs.~4(a) and 4(e).  At $V_{surface}=1.4$ eV, the original surface bands of Bi merge entirely into the bulk conduction along $\overline{\Gamma M}$. A new pair of surface bands appear in the gap. They emerge from the bulk valence band below the Fermi level, but they do not touch the conduction band. The Sb surface bands, on the other hands, always connect to the conduction and valence bands in a topological fashion, no matter how we tune the surface potential; see Figs. 4(f-g). This clearly demonstrates the topological distinction between Bi and Sb.
 
\section{Strain-induced band topology in Bismuth}
So far we have simulated Bi bulk and films with standard first-principles methods and the results exhibit a trivial band structure of Bi in both forms. We have also shown that in thin films the surface band dispersions under surface perturbations can differentiate the topology properties of Bi and Sb. We will show in this section that if pure Bi is topological, it behaves similar to Sb in film forms. Therefore we can use the behavior of surface bands in thin films as an experimental criterion for detecting the true band topology of Bi. The calculated bulk band gap of Bi at $L$ is very small, therefore it is easy to realize a band inversion by applying a lattice strain or effectively reducing the spin-orbit coupling and make the band structure topologically nontrivial. It has been shown theoretically that alloying with Sb, which suppresses the spin-orbit coupling of Bi, can bring Bi to the topological phase. Here we investigate the effect of lattice strain on the band structure of Bi bulk and thin films. In the simulation, we stretch the lattice by 10\% along the $c$-axis, the diagonal of the rhombohedral structure, while keeping the coordinates in perpendicular directions unchanged. This means the interlayer spacing and intralayer buckling parameters are all enlarged by 10\%. The calculated Wannier charge center is plotted in Fig.~5(a) and the result is similar to what we obtained for Sb. The $\mathbb{Z}_{2}$ invariant $\nu$ is 0 and 1 in $k_{z}=0$ and $k_{z}=\frac{\pi}{2}$ planes, respectively, which indicates that Bi under this strain is in the topological phase. The bulk band dispersion of the strained Bi resembles the unstrained one except the electron pocket goes a little deeper at the $L$ point. The gap size at $L$ in this strained case is 0.09 eV. The band structure of a 20L slab shows the same connection pattern as the one in the unstrained case (Fig.~5(b)). When a surface bias is applied to the top surface, the doubly degenerate surface bands split into two copies which are separated spatially on the two surfaces (Fig.~5(c)). The surface projection in Figs.~5(d, e) clearly reveal that in the presence of the surface perturbation, each copy of the surface bands retain the same nontrivial connection to the bulk bands, unlike the case of unstrained Bi (Figs.~3(c,d)).. This is a topological distinction. We can also see this in the semi-infinite slab simulations. The surface bands of strained Bi always maintain a dispersion that connects the conduction and valence bands between $\bar{\Gamma}$ and $\bar{M}$ no matter how strong the surface perturbation is; see Figs. 5(f-j). The strained Bi films behaves exactly in the same way as the Sb films under various surface perturbations. Therefore, the topological invariant of Bi can be detected by tracking the surface band response to surface perturbations in thin films.
{\it We note that the ARPES measurements reported in \cite{Ohtsubo1, Ohtsubo2, Matsuda} examined the surface band dispersion under the bulk limit and strongly suggested a nontrivial band topology of pure Bi. According to our simulations, for bulk samples, Bi band topology can only be detected by showing whether or not there exist in-gap surface states around the L ($\bar{M}$) point. This measurement is very challenging since the L gap is very small ($\sim$ 12 meV), and it has not been achieved experimentally yet due to the limited resolution (typically, a few meV) of the photoemission measurements. Here we provide an alternative way to probe the Bi band topology via studying thin film spectra. The average energy separation between bands in a thin film depends on the hybridization gap, which can be much larger than the intrinsic bulk band gap of Bi as shown in Figs.~5(b-d). The surface band response to surface perturbations can offer decisive evidence for the bulk band topology. This feature in thin films is, in principle, easier to resolve by ARPES because of the much larger energy separation between bands. Therefore, the experimental scheme we proposed in this work overcomes the technical difficulty which exists for bulk Bi samples.}
  
\section{Spin texture and circular dichroism}    
Another interesting feature of Bi and Sb surface states is spin polarization. The topological surface states of Sb must be spin polarized, which is required by the constraint of band topology and time reversal symmetry. Though Bi is not in the topological phase, it possesses rich spin texture on its surface as a consequence of Bychkov-Rashba effect \cite{Rashba} and strong atomic spin-orbit coupling. Circular dichroism angle-resolved photoemission spectroscopy (CD-ARPES) has been proven as a powerful tool for probing complex spin and pseudospin texture on sample surfaces \cite{CD1, CD2, CD3, CD4, CD5, CD6, CD7, CD8}. In an CD-ARPES experiment, circularly polarized photons are used to excite electrons and the resulting dichroic patten, i.e. the difference between the ARPES spectra taken under two circular polarization modes (left-handed mode (LCP) and right-handed mode (RCP)), is closely related to the spin polarization of the surface states or surface resonances. Here we performed CD-ARPES measurements on a Bi(111) film and the experimental geometry is shown in Fig. 6(a). The photon energy is 21~eV. We also calculated the spin texture of Sb and Bi films for a comparison. The Sb(111) surface states are typical Rashba states: the parabolic surface band splits into two spin branches with opposite spin orientations \cite{Rashba}. When the surface bands approach to the bulk band edge, they lose their surface character and spin polarization, and become bulk-like states. The Rashba states are only the surface states on the Sb(111) surface as shown in Fig.~6(b). The Bi(111) surface has, besides the Rashba surface states (SS1) close to the Fermi level, two additional prominent spin features: one is a cross-shaped surface resonance (SR) which overlaps the bulk band in energy; the other is a pair of surface bands (SS2) that reside in a partial bulk band gap as illustrated in Figs.~3(b) and 6(b). The ARPES maps of Bi(111) surface taken under LCP and RCP modes are shown in Fig.~6(c). The asymmetric intensity distribution in the maps leads to a prominent dichroic spectrum (Fig.~6(c)) where the dichroism is defined as $I_{DICH}=(I_{LCP}-I_{RCP})/(I_{LCP}+I_{RCP})$. The maximum dichroism is found to be $\sim0.12$. The dichroic map is in good agreement with the calculated spin texture of Bi (111) surface. All three spin features are resolved in our CD-ARPES experiments, which establishes a clear picture of the spin ingredients of Bi surface that can be utilized in spintronics applications.  We note that CD-ARPES measurements are sensitive to the photon energy as demonstrated in previous experimental works \cite{CD9}. The magnitude and even the sign of dichroism can change with photon energy. A comprehensive interpretation of the dichroic map requires detailed theoretical simulations of the complex photoemission process. We will leave this task for future work.   
 
 \section{Summary}
With the advent of nano-scale electronics, it is of great importance to study electron and spin behaviors on surfaces and thin films. Electronic systems with reduced dimensions can host novel electronic behaviors because of quantum confinement effects. The confinement effect of thin films, as we showed in this work, can create a topological band structure in Bi thin films no matter the bulk material is topological or not. This occurs when the size of the hybridization gap is larger than that of the intrinsic bulk band gap. We can define a critical thickness when the two gaps coincide in size. The true band topology can only be revealed in freestanding films when the film thickness is larger than the critical value. The main difficulty in detecting Bi band topology lies in how to resolve the surface states in the tiny bulk gap at $L$. We demonstrated in our simulations that one can circumvent this difficulty by studying the surface bands in thin films. The response of surface bands in thin films to surface perturbations is an indicator of band topology and, meanwhile, the quantum confinement in thin films generates sizable energy gaps between bands, which facilitate ARPES measurements. The strong surface perturbation can be realized by chemical terminations \cite{Dirac}. The interplay of quantum confinements and surface perturbations generally lead to a variety of band dispersions of surface electrons, which can be of interest for thin film devices. In addition, the complex electronic and spin texture on the Bi surface were mapped out by our CD-ARPES measurements. The results, taken collectively, shed light on the intriguing electronic and spin properties of Bi, especially, in quantum dimensions.  


\newpage

\begin{figure}
\centering
\includegraphics[width=16cm]{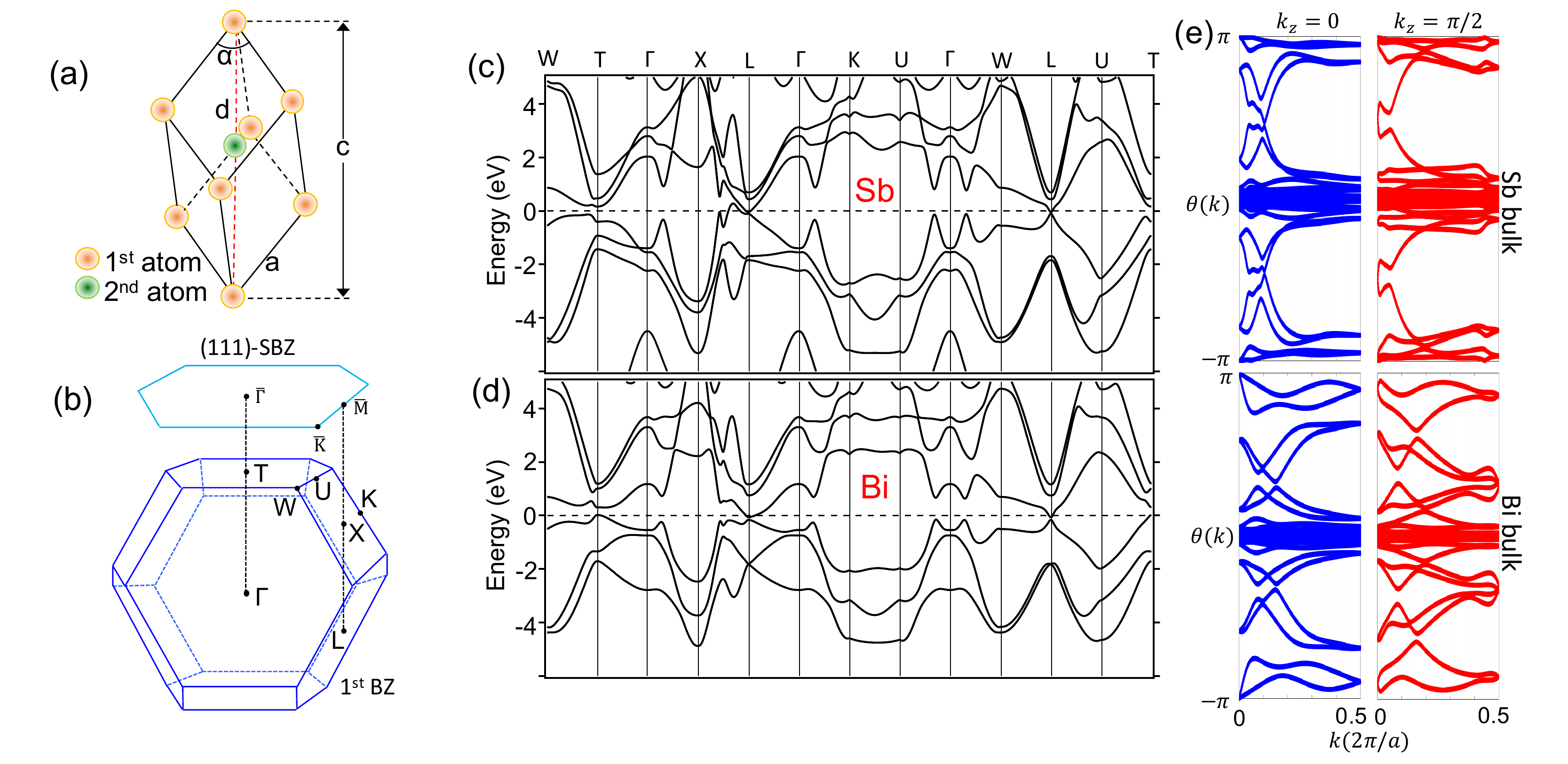}
\caption{(color online). (a) Rhombohedral crystal lattice of Bi and Sb. (b) Bulk Brillouin zone and (111) surface Brillouin zone of Bi and Sb. (c, d) Bulk band structure of Bi and Sb, respectively. (e) Wannier charge center evolution in time-reversal invariant planes of Bi and Sb.}
\end{figure}

\newpage

\begin{figure}
\centering
\includegraphics[width=16cm]{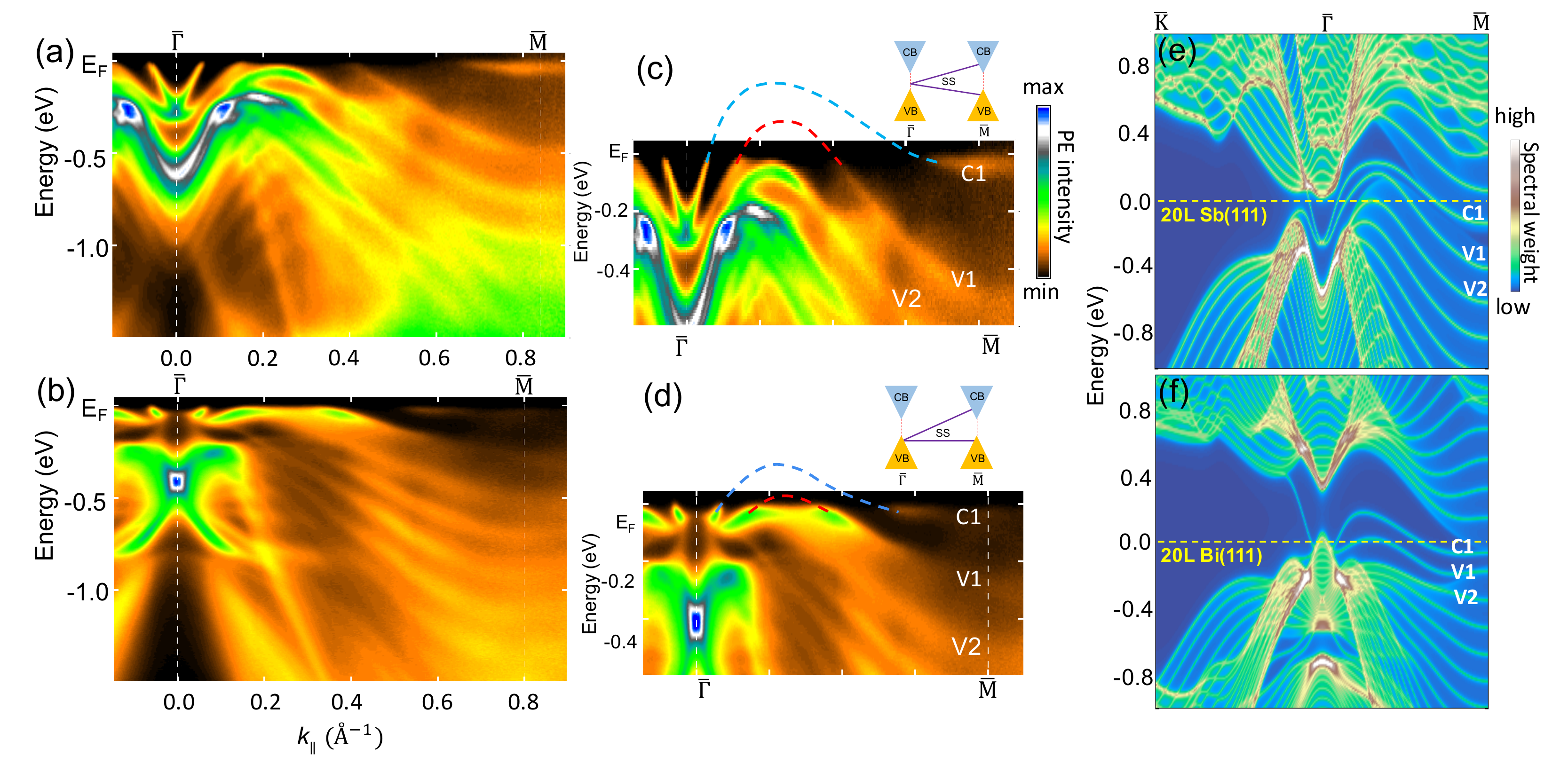}
\caption{(color online). (a) ARPES spectrum along $\overline{\Gamma}-\overline{M}$ taken from a 20L Sb(111) film. (b) Same as (a) but from a 20L Bi(111) film. (c, d) Zoom-in spectra close to the Fermi level from (a,~b), respectively. The dashed lines and schematic drawings show the connection of Rashba surface bands to quantum well conduction and valence subbands. (e) Calculated band structure of 20L Bi(111) and 20L Sb(111) slabs. The shaded areas indicate the projected bulk bands.} 
\end{figure}

\newpage

\begin{figure}
\centering
\includegraphics[width=16cm]{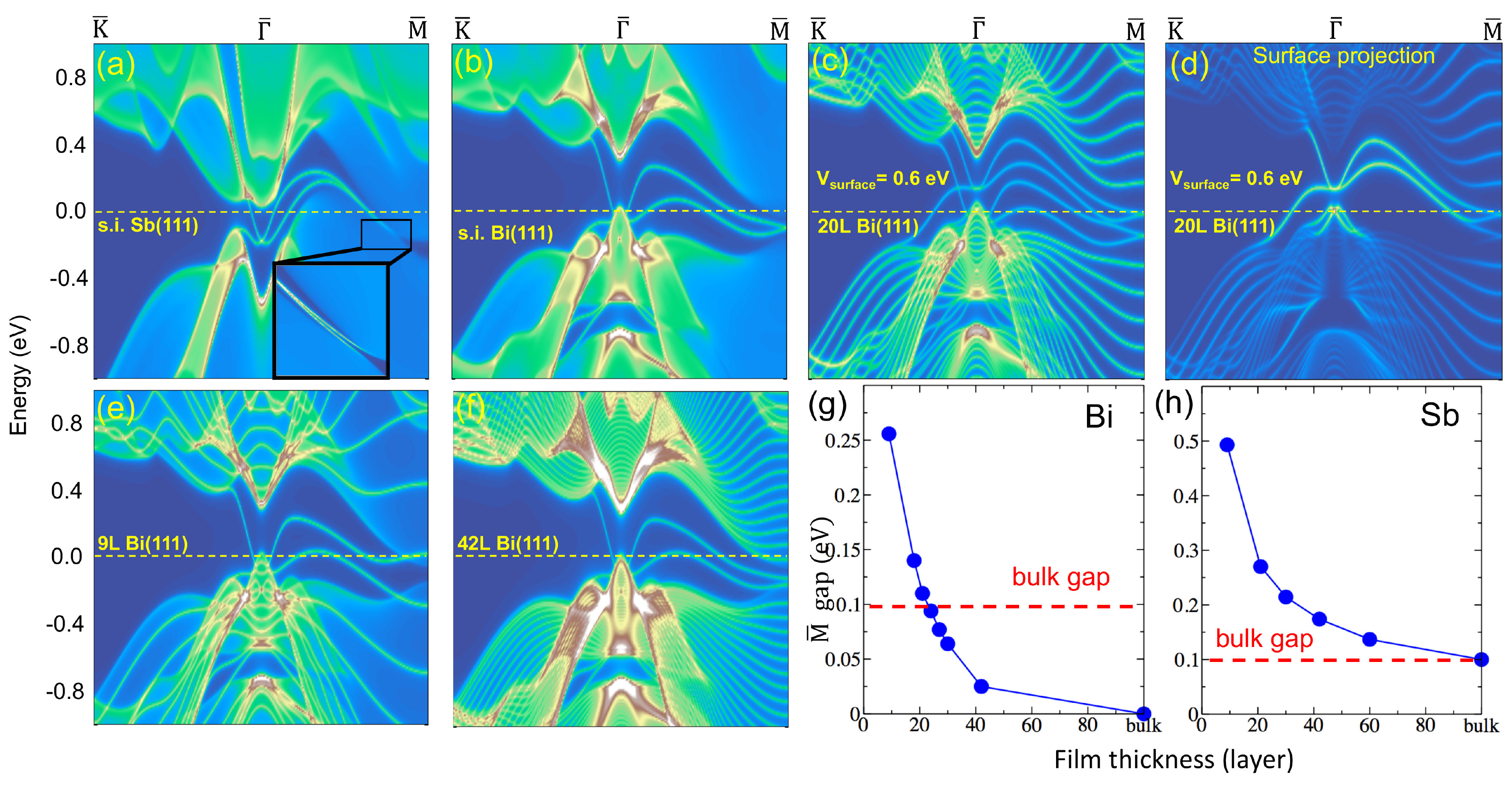}
\caption{(color online). (a, b) Band spectra of semi-infinite Bi and Sb crystals with a (111)-terminated surface. (c) Calculated band structure of 20L Bi(111) with a surface potential 0.6 eV on the top layer. (d) Same as (c) but weighted with the charge density at the top layer. (e,f) Band structure of 9L and 42L Bi(111) slabs. (g) The size of the band gap between the two surface bands at $\overline{M}$ as a function of Bi film thickness. (h) Same as (g), but for Sb films.}
\end{figure}

\newpage

\begin{figure}
\centering
\includegraphics[width=16cm]{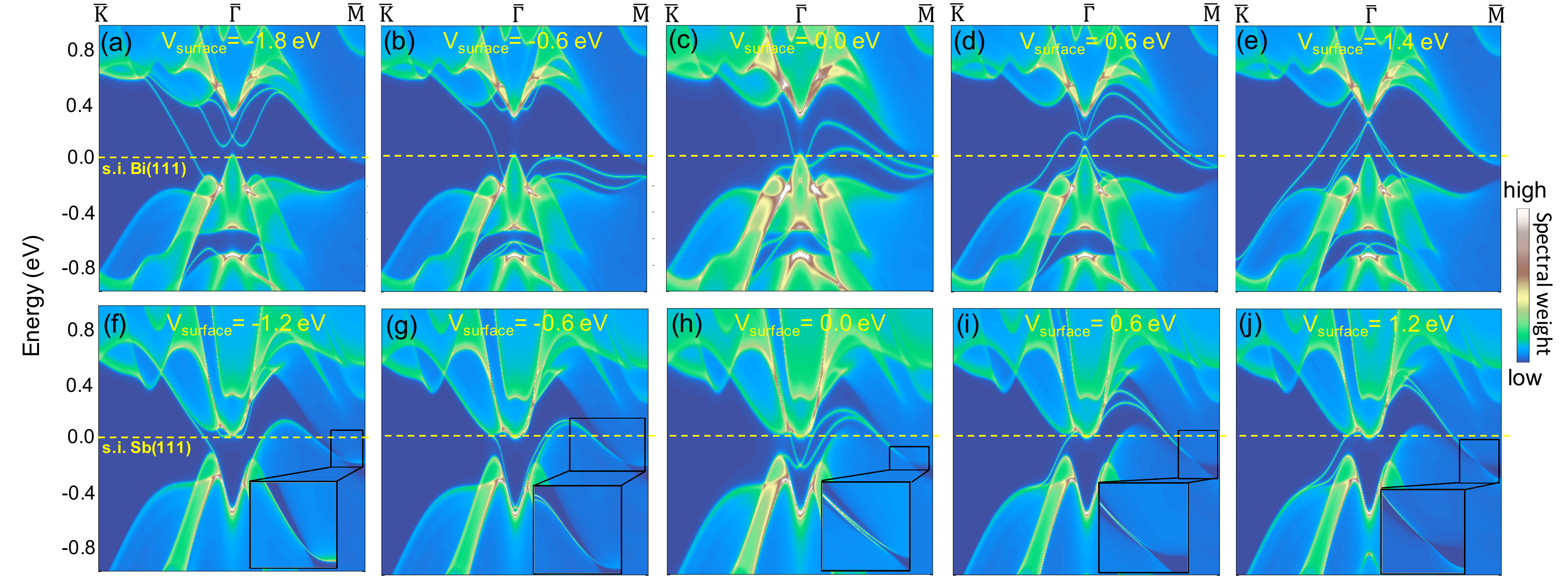}
\caption{(a-e) Band spectrum of semi-infinite Bi under different surface potential bias. (f-j) Same as (a-e) but for Sb. }
\end{figure}

\newpage

\begin{figure}
\centering
\includegraphics[width=16cm]{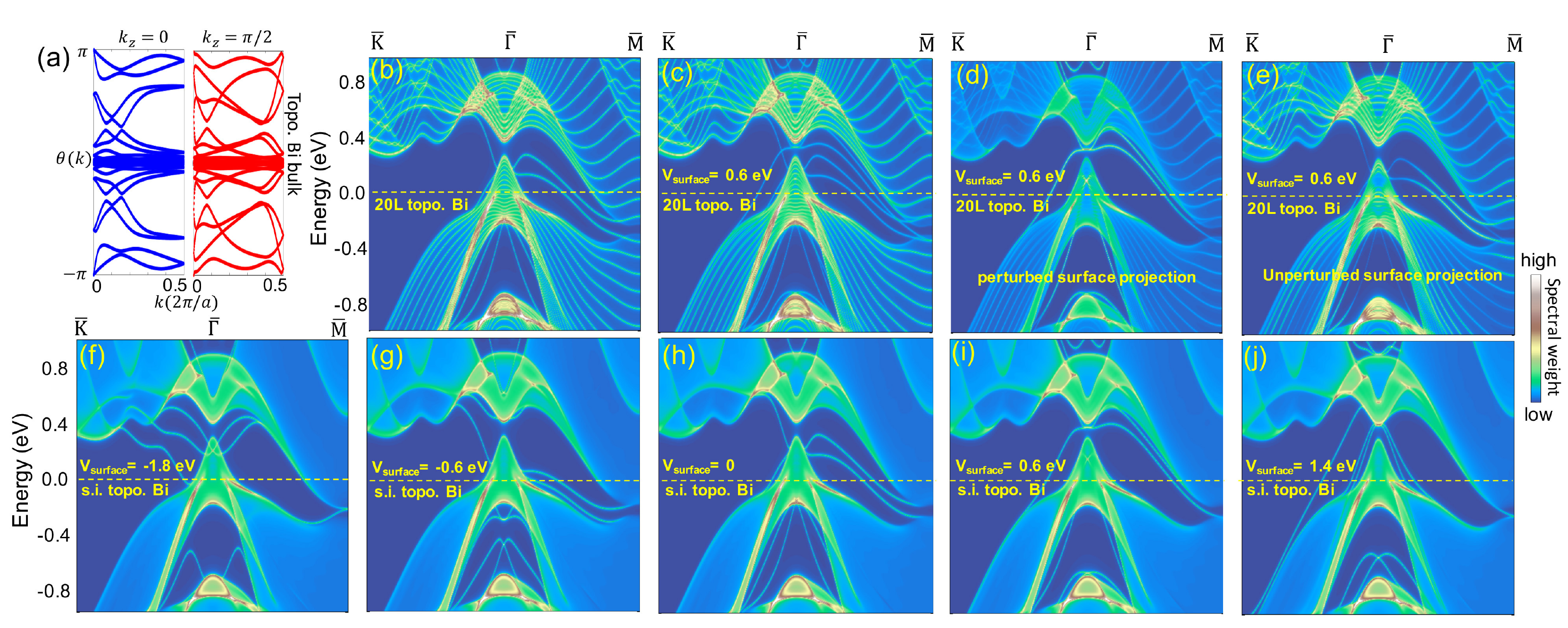}
\caption{(color online). (a) Wannier charge center evolution in the time-reversal invariant planes of strained Bi. The atomic coordinate along c-axis is stretched by 10\%. The strained Bi is topologically nontrivial.  (b) Calculated band structure of a 20L strained Bi(111) film. (c) Band structure of a 20L Bi(111) with a surface potential 0.6 eV on the top layer. (d, e) Same as (c) but weighted with the charge density at the perturbed top layer and the unperturbed bottom layer, respectively. (f-j) Band spectrum of semi-infinite strained Bi under different surface potential bias.}
\end{figure}

\newpage

\begin{figure}
\centering
\includegraphics[width=16cm]{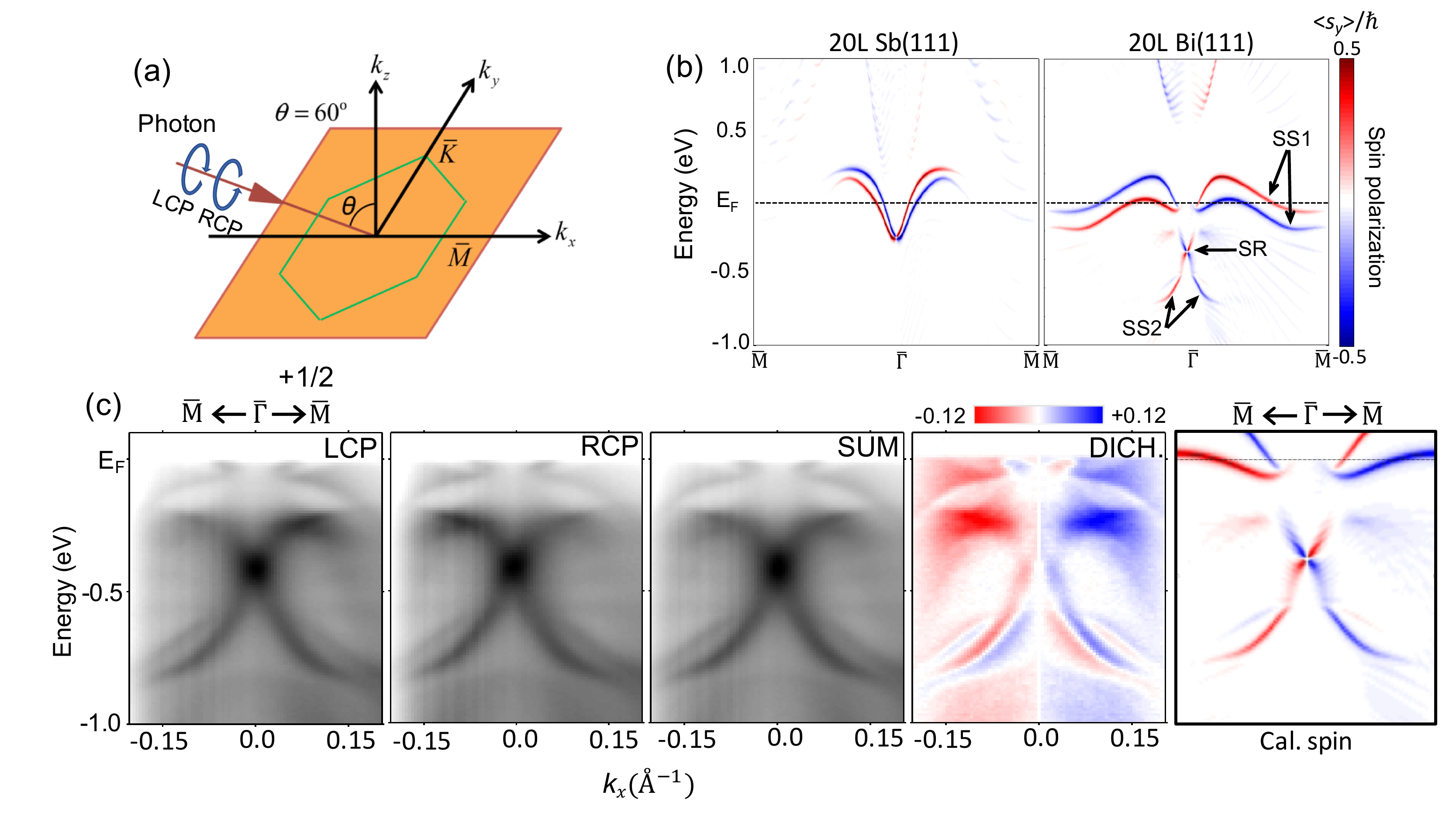}
\caption{(color online). (a) Experimental geometry of the circular ARPES measurement. (b) Calculated spin texture of 20L Sb(111) and 20L Bi(111) films. (c) Circular dichroism taken from a 20L Bi(111) film. The panels from left to right show the APRES spectra taken with light and right-handed circularly polarized (LCP and RCP) photons, the summation of LCP and RCP data, the measured circular dichroism, and the calculated spin texture.}
\end{figure}

\end{document}